\documentclass[twoside]{article}
\usepackage{fleqn,espcrc2}

\usepackage{graphicx}

\title{Large Effects of CP- and T-Violation in $K^0$ Decay}

\author{L. M. Sehgal
        \address{Institute of Theoretical Physics (E), RWTH Aachen \\ 52056 Aachen, Germany}
        \thanks{Talk at {\it International Conference on CP Violation Physics}, 18-22 September 2000, Ferrara, Italy.}
        }
       
\begin{document}

\begin{abstract}
The $\epsilon$-impurity of the $K_L$ wave-function gives rise to a huge
$CP$-violating effect in the decay $K_L \to \pi^+ \pi^- \gamma$ which is hidden
in the polarization state (Stokes vector) of the photon. One component of the
Stokes vector is $CP$-odd and $T$-even, and may be identified with circular
polarization. Another component is $CP$-odd and $T$-odd
(``oblique polarization'')
and reveals itself as a large asymmetry ($14\%$) in the decay
$K_L \to \pi^+ \pi^- e^+ e^-$. Striking time-dependent effects are predicted
in the angular distribution of the $\pi^+ \pi^- e^+ e^-$ system emanating
from an initial $K^0$ or $\overline{K^0}$ state.
\vspace{1pc}
\end{abstract}

\maketitle

It is not customary to use the word "large" in association with $CP$-violating
effects in $K^0$ decays. Nevertheless, a large effect has been observed in
the decay $K_L \to \pi^+ \pi^- e^+ e^-$~\cite{KTeV:Cox,NA48}, in agreement
with theoretical predictions~\cite{Sehgal:Wanninger,Heiliger:Sehgal}. In
this talk, I explain the origin of this effect, and some of its ramifications.

\section{$CP$- and $T$-Violation in $K_L \to \pi^+ \pi^- \gamma$}

The decay $K_S \to \pi^+ \pi^- \gamma$ is known to be well-described by pure
bremsstrahlung. By contrast, the branching ratio and the photon energy spectrum
of $K_L \to \pi^+ \pi^- \gamma$ require a mixture of bremsstrahlung and direct
$M1$ emission~\cite{Ramberg,Data}. A simple ansatz for the matrix elements 
is~\cite{Sehgal:Leusen99}
\begin{eqnarray}
\lefteqn{{\cal M}(K_{S,L} \to \pi^+ \pi^- \gamma)  =} \nonumber\\
\lefteqn{\hspace{10pt} \frac{e |f_S|}{M^4_K} \left\{ E_{S,L}(\omega, \cos \theta) \left[ \epsilon \cdot
p_+ k \cdot p_- - \epsilon \cdot p_- k \cdot p_+ \right] \right.} \nonumber\\
\lefteqn{\hspace{10pt} + \left. M_{S,L}(\omega, \cos \theta) \epsilon_{\mu \nu \rho \sigma} \epsilon^{\mu}
k^{\nu} p^{\rho}_+ p^{\sigma}_- \right\},}
\end{eqnarray}
where
\begin{eqnarray}
E_S & = & \left( \frac{2 M_K}{\omega} \right)^2
\frac{e^{i \delta_0(s = M^2_K)}}{1 - \beta^2 \cos^2 \theta}, \hspace{10pt} M_S = 0,
\nonumber\\
E_L & = & \left( \frac{2 M_K}{\omega} \right)^2
\frac{\eta_{+-} e^{i \delta_0(s = M^2_K)}}{1 - \beta^2 \cos^2 \theta},
\nonumber\\
M_L & = & i(0.76)e^{i\delta_1(s)}.
\end{eqnarray}
Here $\omega$ is the photon energy in the $K_L$ rest frame, $\theta$ the angle
between $\pi^+$ and $\gamma$ in the $\pi^+ \pi^-$ c.m. frame, and
$\beta = \sqrt{1- \frac{4 m^2_{\pi}}{s}}$, $s$ being the $\pi^+ \pi^-$ invariant
mass. The coefficient $0.76$ in $M_L$ is determined from the empirical strength
of direct emission in $K_L \to \pi^+ \pi^- \gamma$~\cite{Sehgal:Wanninger}.
The phase factor $e^{i \delta_0(M^2_K)}$ in the bremsstrahlung amplitudes $E_{L,S}$
is dictated by the Low theorem, while the factor $ie^{i \delta_1(s)}$ in $M_L$ is
determined by $CPT$ invariance and the Watson theorem. The important feature of
the $K_L$ amplitude is that the bremsstrahlung component $E_L$, proportional to
$\eta_{+-}$, is enhanced by the factor $(2 M_K / \omega)^2$, making it
comparable to the direct emission amplitude $M_L$. The interference of the electric
multipoles ($CP = + 1$) with the magnetic multipole ($CP = -1$) opens the way to
large $CP$-violating observables. Such interference effects vanish, however, if
one sums over the photon polarization. Thus $CP$ violation is encrypted in the
polarization state of the photon.

The polarization state of the photon can be described by the Stokes vector
$\vec{S} = (S_1,S_2,S_3)$ whose components are (dropping the subscript $L$ in
$E_L$, $M_L$)
\begin{eqnarray}
S_1 & = & \frac{2 {\rm Re} E^* M}{|E|^2 + |M|^2}, \nonumber\\
S_2 & = & \frac{2 {\rm Im} E^* M}{|E|^2 + |M|^2}, \nonumber\\
S_3 & = & \frac{|E|^2 - |M|^2}{|E|^2 + |M|^2}.
\end{eqnarray}

\begin{figure}[htb]
\makebox[75mm]{
\resizebox{75mm}{47mm}
{\includegraphics*[19mm,26mm][280mm,188mm]{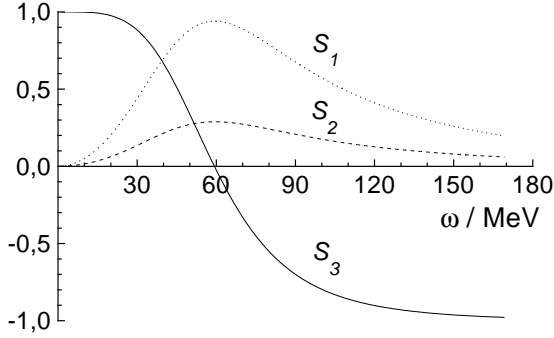}}
}
\vspace{-40pt}
\caption{Stokes parameters of the photon in $K_L \to \pi^+ \pi^- \gamma$}
\label{S1:S2:S3}
\end{figure}

These are plotted in Fig.~\ref{S1:S2:S3} as a function of the photon energy. The
components $S_1$ and $S_2$ are $CP$-violating observables, and are remarkably
large, considering that they originate in the small parameter
$\epsilon \approx \eta_{+-}$. To see the physical meaning of these parameters,
we choose a frame in which $\vec{k} = (0,0,\omega)$ and
$\vec{n}_{\pi} = (\vec{p}_+ \times \vec{p}_-) / |\vec{p}_+ \times \vec{p}_-| = (1,0,0)$.
The parameter $S_2$ is then recognized as the {\it circular polarization}, i.e.
\begin{equation}
S_2 = [d\Gamma (L) - d\Gamma (R)] / [d\Gamma (L) + d\Gamma (R)]
\end{equation}
where $L$ and $R$ refer to the polarization vectors
$\vec{\epsilon}_{R,L} = (1, \pm i, 0) / \sqrt{2}$. This is a $CP$-odd but $T$-even
observable, and vanishes in the hermitian limit $\delta_0 = \delta_1 = 0$,
${\rm arg} \eta_{+-} = \pi /2$~\cite{Sehgal:Leusen99}. To understand the
significance of $S_1$, we consider the linear polarization vector
$\vec{\epsilon} = (\cos \phi, \sin \phi, 0)$, where $\phi$ is the polarization
direction relative to $\vec{n}_{\pi}$, the normal to the decay plane. One then
discovers that $S_1$ is the ``{\it oblique polarization}'', defined as the
difference between the decay rates for $\phi = 45^{\circ}$ and $\phi = 135^{\circ}$:
\begin{equation}
S_1 = \frac{d\Gamma (45^{\circ}) - d\Gamma (135^{\circ})}
{d\Gamma (45^{\circ}) + d\Gamma (135^{\circ})}.
\end{equation}
More generally, the decay rate as a function of $\phi$ is
\begin{equation}
\frac{d\Gamma}{d\phi} \sim 1 - \left[ S_3 \cos 2\phi + S_1 \sin 2\phi \right].
\end{equation}

\begin{figure}[htb!]
\makebox[75mm]{
\resizebox{75mm}{56mm}
{\includegraphics*[12mm,13mm][262mm,200mm]{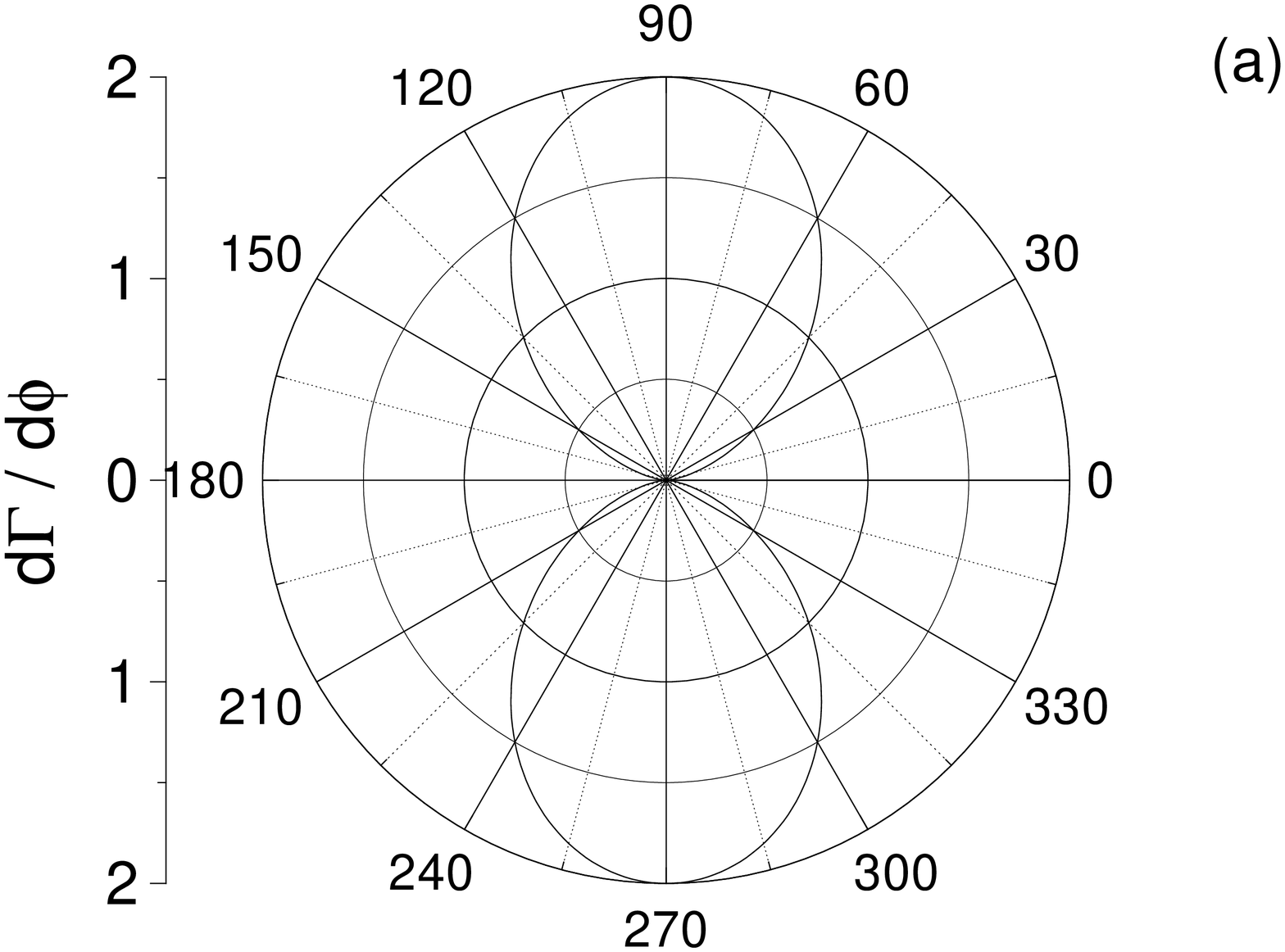}}
}
\makebox[75mm]{
\resizebox{75mm}{56mm}
{\includegraphics*[12mm,13mm][262mm,200mm]{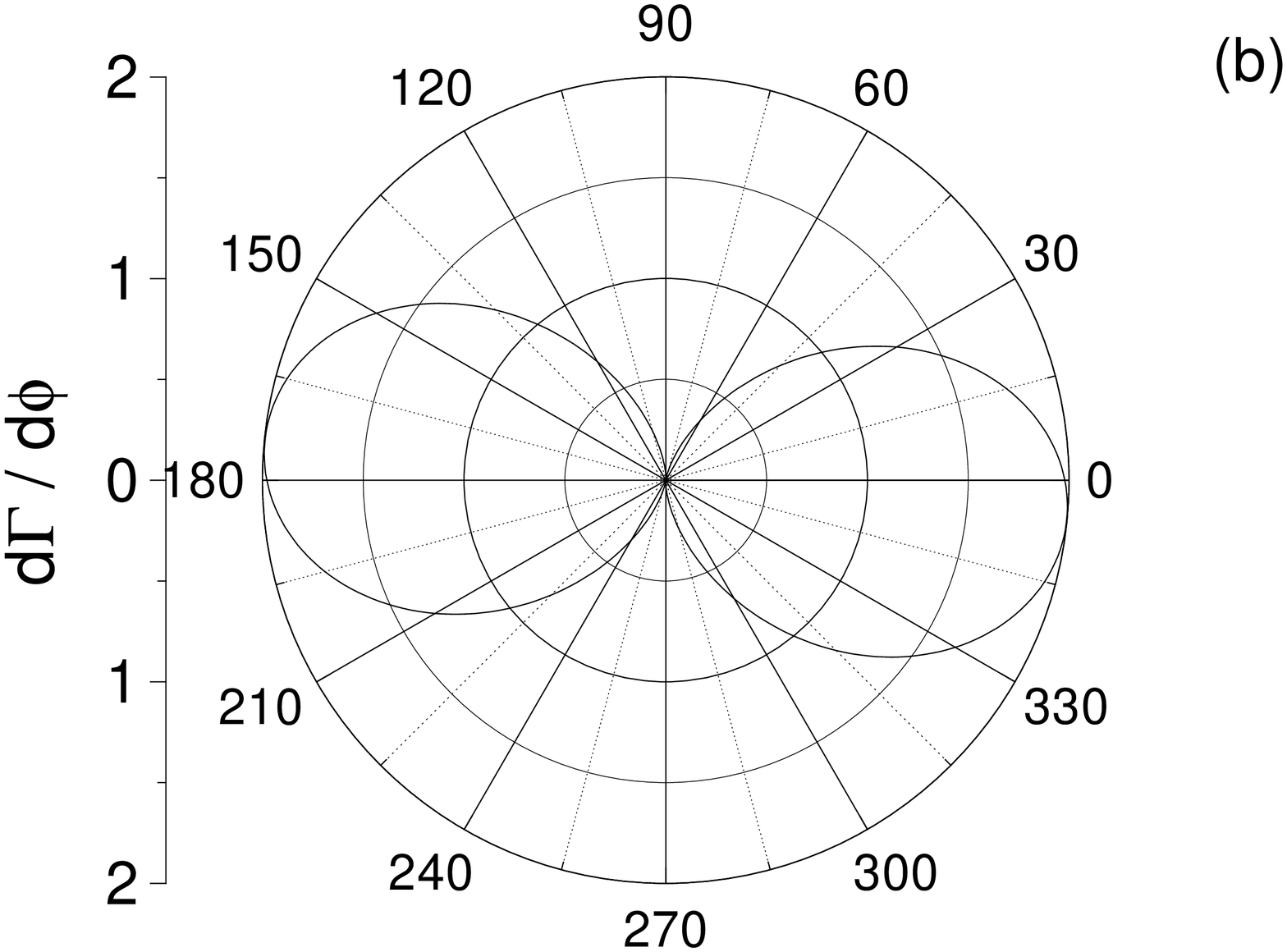}}
}
\makebox[75mm]{
\resizebox{75mm}{56mm}
{\includegraphics*[12mm,13mm][262mm,200mm]{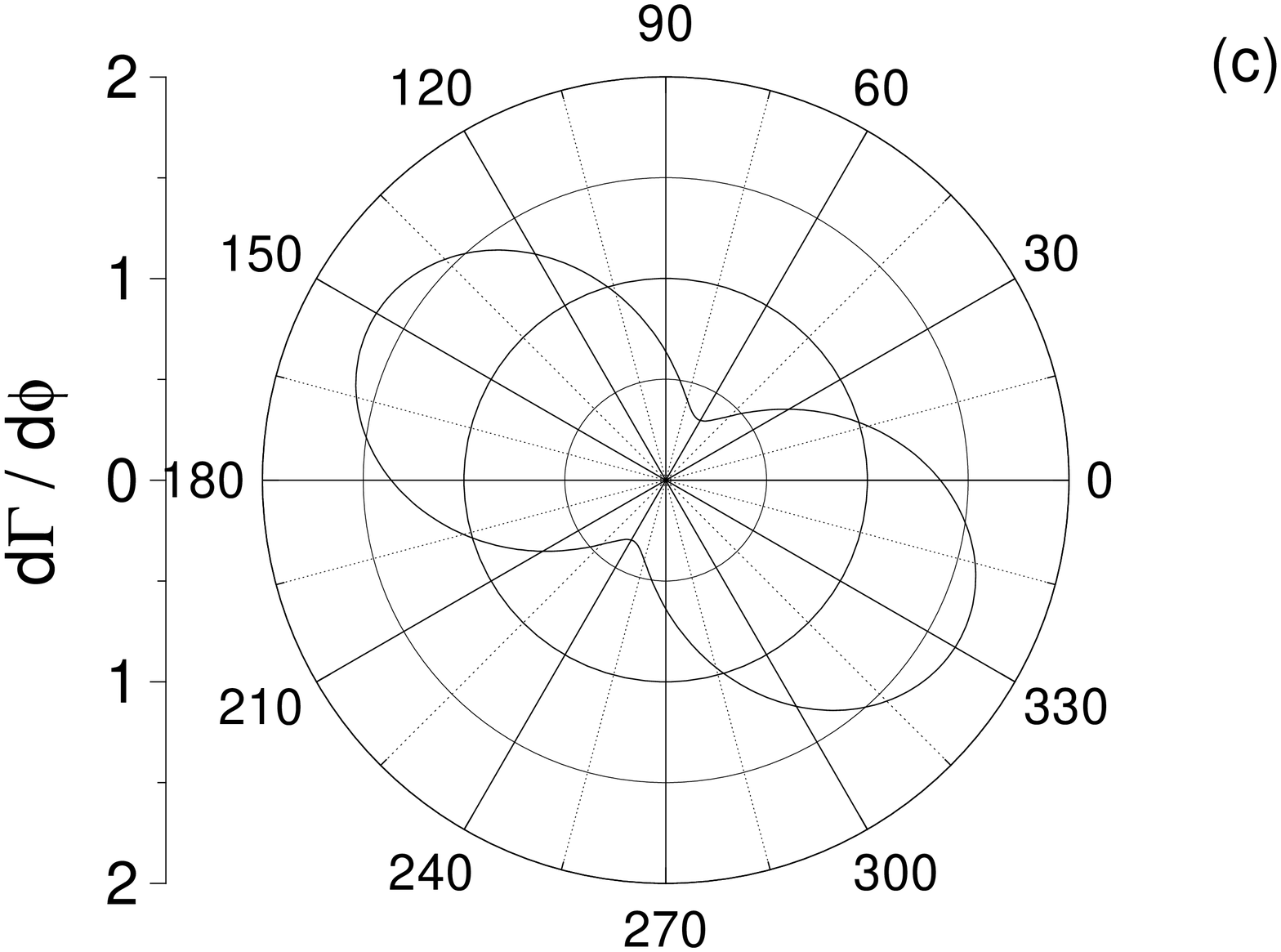}}
}
\vspace{-40pt}
\caption{Distribution of $K_L \to \pi^+ \pi^- \gamma$ in angle $\phi$ between polarization
vector $\vec{\epsilon}$ and normal to decay plane. (a) $\omega \to 0$, (b) $\omega \to 170\,
 MeV$, (c) average over $20\, MeV < \omega < 170 \, MeV$.}
\label{RateGamma}
\end{figure}

Fig.~\ref{RateGamma} illustrates how this pattern changes from an electric
distribution $d\Gamma / d\phi \sim \sin^2 \phi$ at low $\omega$ to a magnetic
distribution $d\Gamma / d\phi \sim \cos^2 \phi$ as the photon energy increases.
The presence of the parameter $S_1$ produces a tilted pattern in which there is
an asymmetry between the quadrants I+III compared to II+IV. It is this tilt that
signals a violation of $CP$. By examining the behaviour of $\vec{k}$, $\vec{\epsilon}$
and $\vec{n}_{\pi}$ under $CP$ and $T$, we find that $\sin 2 \phi$ transforms as
$CP = -$, $T = -$. Thus the oblique polarization $S_1$ is a $CP$-odd, $T$-odd
observable~\cite{Sehgal:Leusen99}. Unlike the parameter $S_2$, the parameter $S_1$ survives
in the hermitian limit. In this sense, the $T$-odd property of $S_1$ is not an artifact 
of dynamical phases, but rather an example of $T$-violation accompanying $CP$-violation
in a $CPT$ invariant theory~\cite{Sehgal:Leusen99}. The study of the Dalitz pair reaction
$K_L \to \pi^+ \pi^- e^+ e^-$ may be viewed as an attempt to detect the oblique
polarization of the photon in $K_L \to \pi^+ \pi^- \gamma$, using the plane of the
$e^+ e^-$ pair as an analyser.

\section{$CP$- and $T$-Violation in $K_L \to \pi^+ \pi^- e^+ e^-$}

The matrix element of the reaction $K_L \to \pi^+ \pi^- e^+ e^-$ can be written
as~\cite{Sehgal:Wanninger,Heiliger:Sehgal}
\begin{equation}
{\cal M} = {\cal M}_{br} + {\cal M}_{mag} + {\cal M}_{CR} + {\cal M}_{SD}
\label{DefM}
\end{equation}
where the ${\cal M}_{br}$ and ${\cal M}_{mag}$ are associated with the
bremsstrahlung and $M1$ components of the radiative amplitude. The term
${\cal M}_{CR}$ denotes a ``charge radius'' contribution, corresponding to
$\pi^+ \pi^-$ emission in an $s$-wave, not possible for the real radiative process
$K_L \to \pi^+ \pi^- \gamma$. The term ${\cal M}_{SD}$ represents the contribution
of the short-distance interaction $s\overline{d} \to e^+ e^-$. Estimates
in~\cite{Sehgal:Wanninger,Heiliger:Sehgal} showed that the amplitude is
dominated by the first two terms in Eq.(\ref{DefM}), as is now borne out by the
data~\cite{KTeV:Cox,NA48}. The differential decay rate may be calculated in the
form
\begin{eqnarray}
\lefteqn{d\Gamma =} \\
\lefteqn{\hspace{10pt} I(s_{\pi}, s_l, \cos \theta_l, \cos \theta_{\pi}, \phi)
ds_{\pi} ds_l d\cos \theta_l d\cos \theta_{\pi} d\phi} \nonumber
\end{eqnarray}
where $s_{\pi} (s_l)$ is the invariant mass of the pion (lepton) pair and
$\theta_{\pi} (\theta_l)$ is the angle of the $\pi^+ (l^+)$ in the
$\pi^+ \pi^- (l^+ l^-)$ rest frame, relative to the dilepton (dipion) direction.
For the purpose of detecting the oblique polarization in $K_L \to \pi^+ \pi^- \gamma$,
the relevant variable is $\phi$, the angle between the normals to the $\pi^+ \pi^-$
and $l^+ l^-$ planes. Defining the unit vectors
\begin{eqnarray}
\vec{n}_{\pi} & = &
\frac{\vec{p}_+ \times \vec{p}_-}{|\vec{p}_+ \times \vec{p}_-|}, \nonumber\\
\vec{n}_{l} & = &
\frac{\vec{k}_+ \times \vec{k}_-}{|\vec{k}_+ \times \vec{k}_-|}, \nonumber\\
\vec{z} & = &
\frac{\vec{p}_+ + \vec{p}_-}{|\vec{p}_+ + \vec{p}_-|},
\end{eqnarray}
we have
\begin{eqnarray}
\sin \phi & = & \vec{n}_{\pi} \times \vec{n}_{l} \cdot \vec{z}
\hspace{10 pt} (CP = -, T = -)\nonumber\\
\cos \phi & = & \vec{n}_{l} \cdot \vec{n}_{\pi}
\hspace{28pt} (CP = +, T = +).
\end{eqnarray}

\begin{figure}[htb]
\makebox[75mm]{
\resizebox{75mm}{56mm}
{\includegraphics*[12mm,13mm][262mm,200mm]{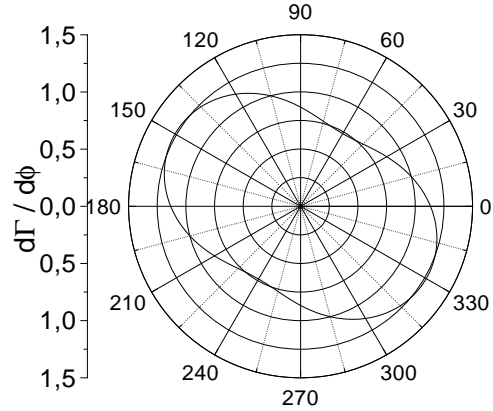}}
}
\vspace{-40pt}
\caption{Distribution of $K_L \to \pi^+ \pi^- e^+ e^-$ in angle $\phi$ between
$\pi^+ \pi^-$ and $e^+ e^-$ planes.}
\label{RateEE}
\end{figure}

Integrating over all variables other than $\phi$, one obtains~\cite{Sehgal:Leusen99}
\begin{equation}
\frac{d\Gamma}{d\phi} \sim 1 - \left( \Sigma_3 \cos 2 \phi + \Sigma_1 \sin 2 \phi \right)
\end{equation}
with $\Sigma_3 = - 0.133$ and $\Sigma_1 = 0.23$. This distribution is plotted
in Fig.~\ref{RateEE}, and shows clearly the $CP$- and $T$-violating ``tilt'' similar
to the oblique polarisation in Fig.~\ref{RateGamma}. The KTeV and NA48 experiments
confirm the predicted $\phi$-distribution~\cite{KTeV:Cox,NA48}, and also the
integrated asymmetry
\begin{eqnarray}
{\cal A}_{\phi} & = &
\frac{\left( \int_{0}^{\pi/2} - \int_{\pi/2}^{\pi} + \int_{\pi}^{3\pi/2}
- \int_{3\pi/2}^{2\pi}\right)\frac{d\Gamma}{d\phi}}{\left( \int_{0}^{\pi/2}
+ \int_{\pi/2}^{\pi} + \int_{\pi}^{3\pi/2} + \int_{3\pi/2}^{2\pi}\right)
\frac{d\Gamma}{d\phi}} \nonumber\\
& = & - 15\% \sin (\delta_0- \delta_1 + \varphi_{+-}) = - 14\%. \label{Asymmetry}
\end{eqnarray}

\section{Time-Evolution of the Decay Spectrum in
$K^0 (\overline{K^0}) \to \pi^+ \pi^- e^+ e^-$~\cite{Sehgal:Leusen00}}

Consider the decay $K^0 (\overline{K^0}) \to \pi^+ \pi^- \gamma$ of a state that is
prepared as an eigenstate of strangeness $+1(-1)$. The decay amplitude at a
subsequent time $t$ can be expressed in terms of the amplitudes $E_{L,S}$, $M_{L,S}$
as follows:

\begin{figure}[htb]
\makebox[75mm]{
\resizebox{75mm}{53mm}
{\includegraphics*{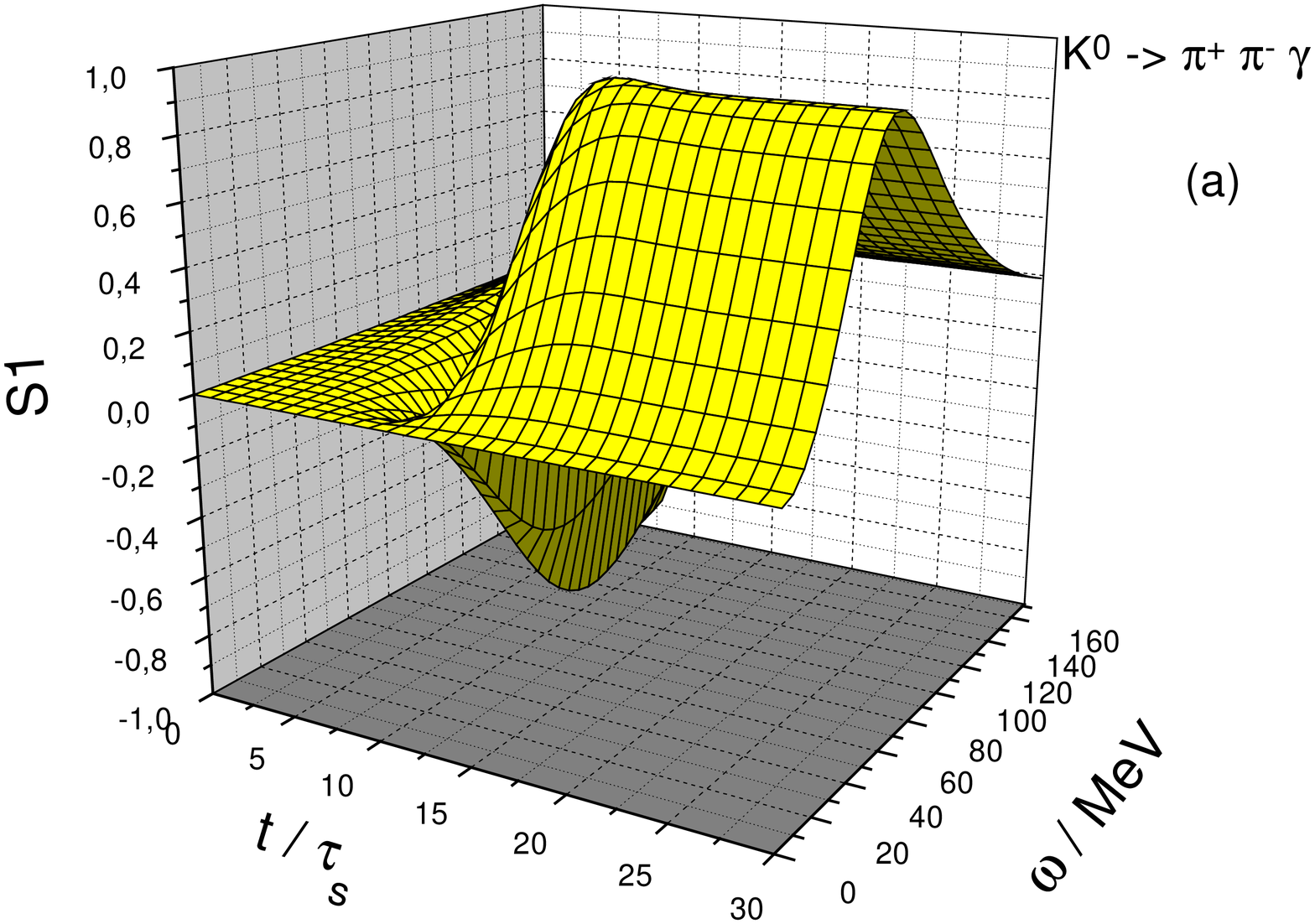}}
}
\makebox[75mm]{
\resizebox{75mm}{56mm}
{\includegraphics*{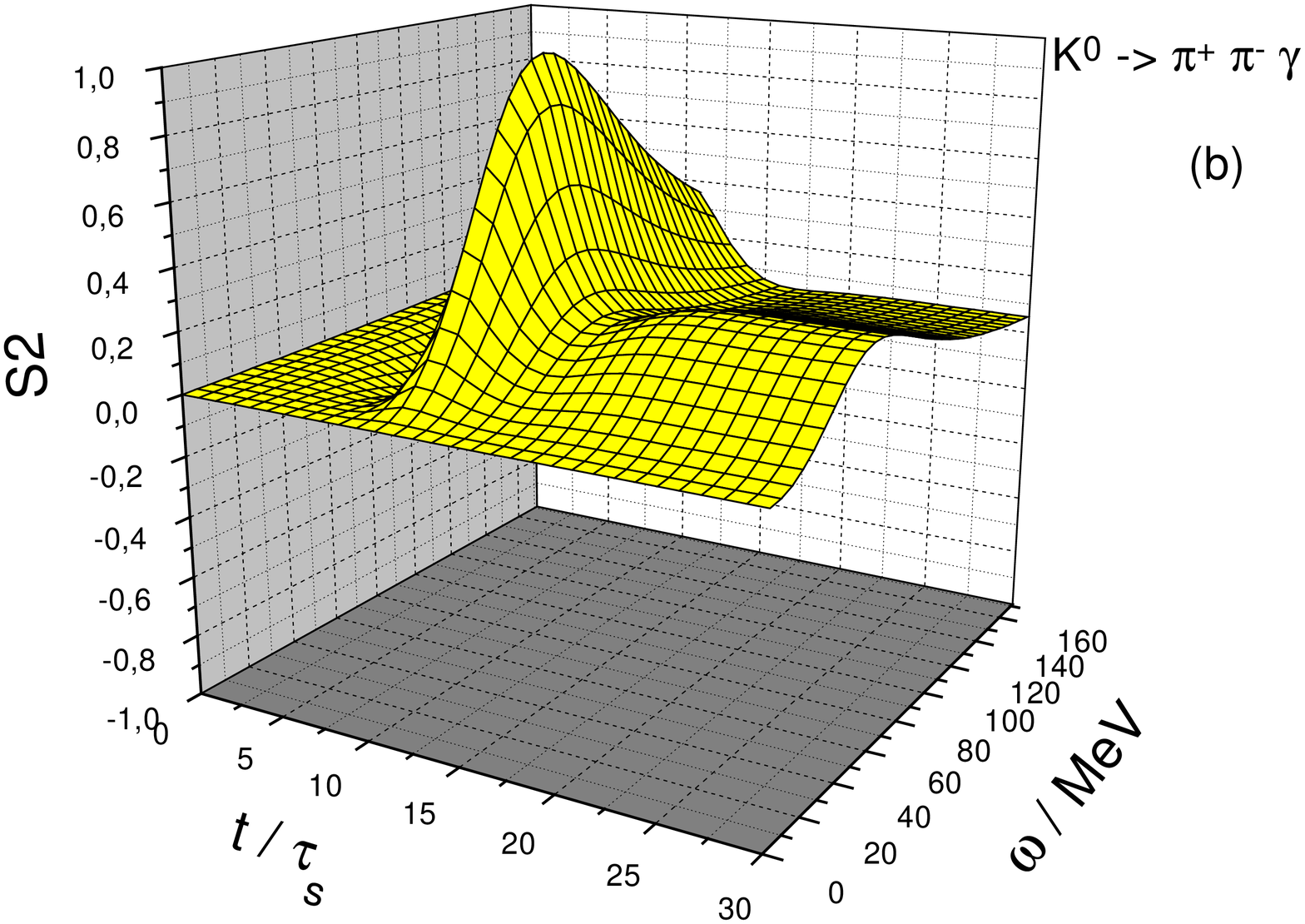}}
}
\vspace{-40pt}
\caption{Components $S_1$ and $S_2$ of the Stokes vector of the photon as a
function of photon energy and time for the decay $K^0 \to \pi^+ \pi^- \gamma$.}
\label{S1t:S2ta}
\end{figure}

\begin{eqnarray}
\lefteqn{{\cal M}(K^0(t) \to \pi^+ \pi^- \gamma) \sim} \nonumber\\
& & \left\{ E(t, \omega, \cos \theta) \left[ \epsilon \cdot
p_+ k \cdot p_- - \epsilon \cdot p_- k \cdot p_+ \right] \right. \nonumber\\
& & \left. + M(t, \omega, \cos \theta) \epsilon_{\mu \nu \rho \sigma} \epsilon^{\mu}
k^{\nu} p^{\rho}_+ p^{\sigma}_- \right\}, \nonumber\\
&&\nonumber\\
\lefteqn{{\cal M}(\overline{K^0}(t) \to \pi^+ \pi^- \gamma) \sim} \nonumber\\
& & \left\{ \overline{E}(t, \omega, \cos \theta) \left[ \epsilon \cdot
p_+ k \cdot p_- - \epsilon \cdot p_- k \cdot p_+ \right] \right. \nonumber\\
& & \left. + \overline{M}(t, \omega, \cos \theta) \epsilon_{\mu \nu \rho \sigma}
\epsilon^{\mu} k^{\nu} p^{\rho}_+ p^{\sigma}_- \right\} \label{AmpK0}
\end{eqnarray}
where
\begin{eqnarray}
E & = & e^{-i\lambda_S t}E_S(\omega, \cos \theta) +
e^{-i\lambda_L t}E_L(\omega, \cos \theta), \nonumber\\
M & = & e^{-i\lambda_L t}M_L(\omega, \cos \theta), \nonumber\\
\overline{E} & = & e^{-i\lambda_S t}E_S(\omega, \cos \theta) -
e^{-i\lambda_L t}E_L(\omega, \cos \theta), \nonumber\\
\overline{M} & = & - e^{-i\lambda_L t}M_L(\omega, \cos \theta)
\end{eqnarray}
with $\lambda_{L,S} = m_{L,S} - \frac{i}{2} \Gamma_{L,S}$.

\begin{figure}[htb]
\makebox[75mm]{
\resizebox{75mm}{53mm}
{\includegraphics*{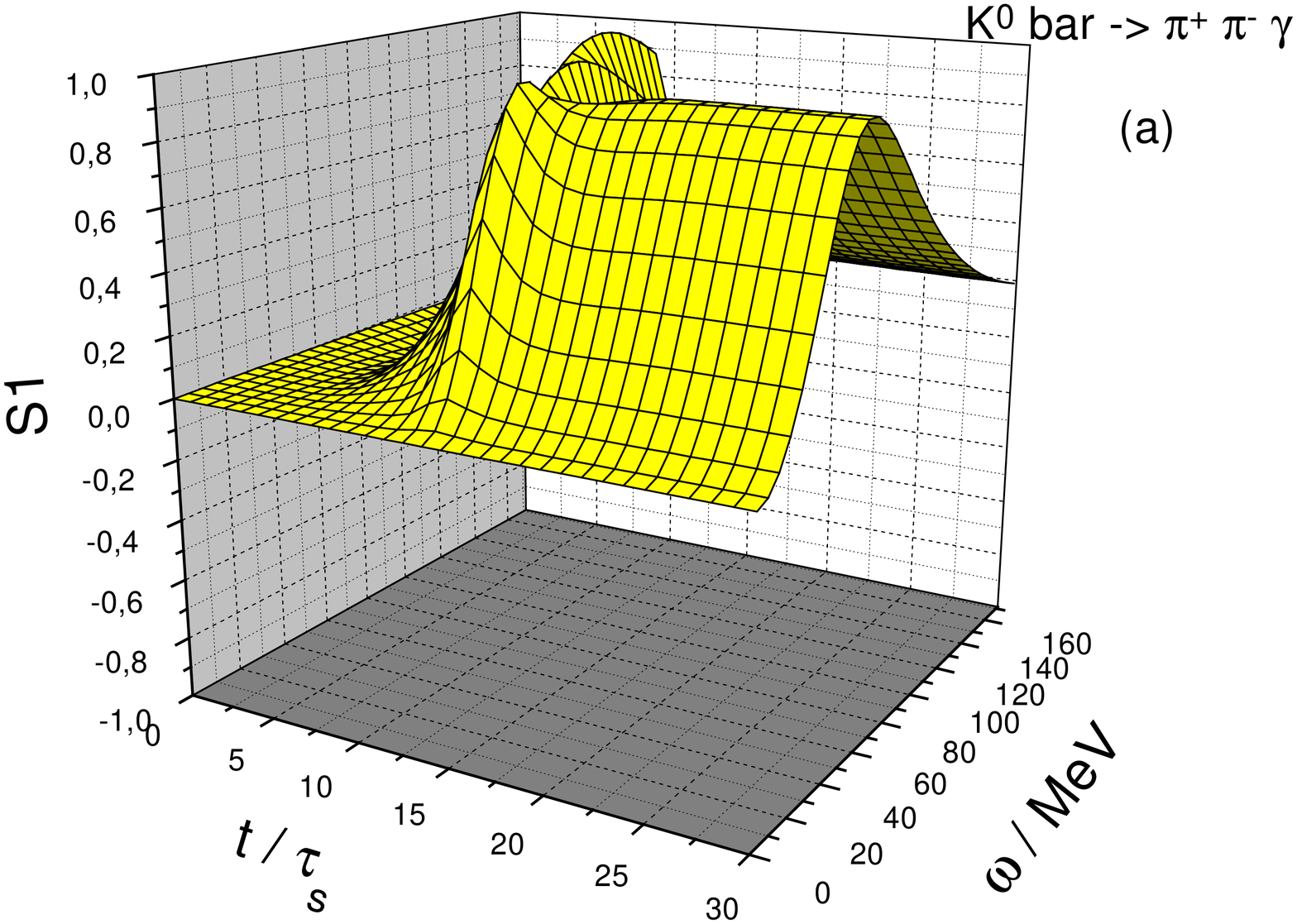}}
}
\makebox[75mm]{
\resizebox{75mm}{56mm}
{\includegraphics*{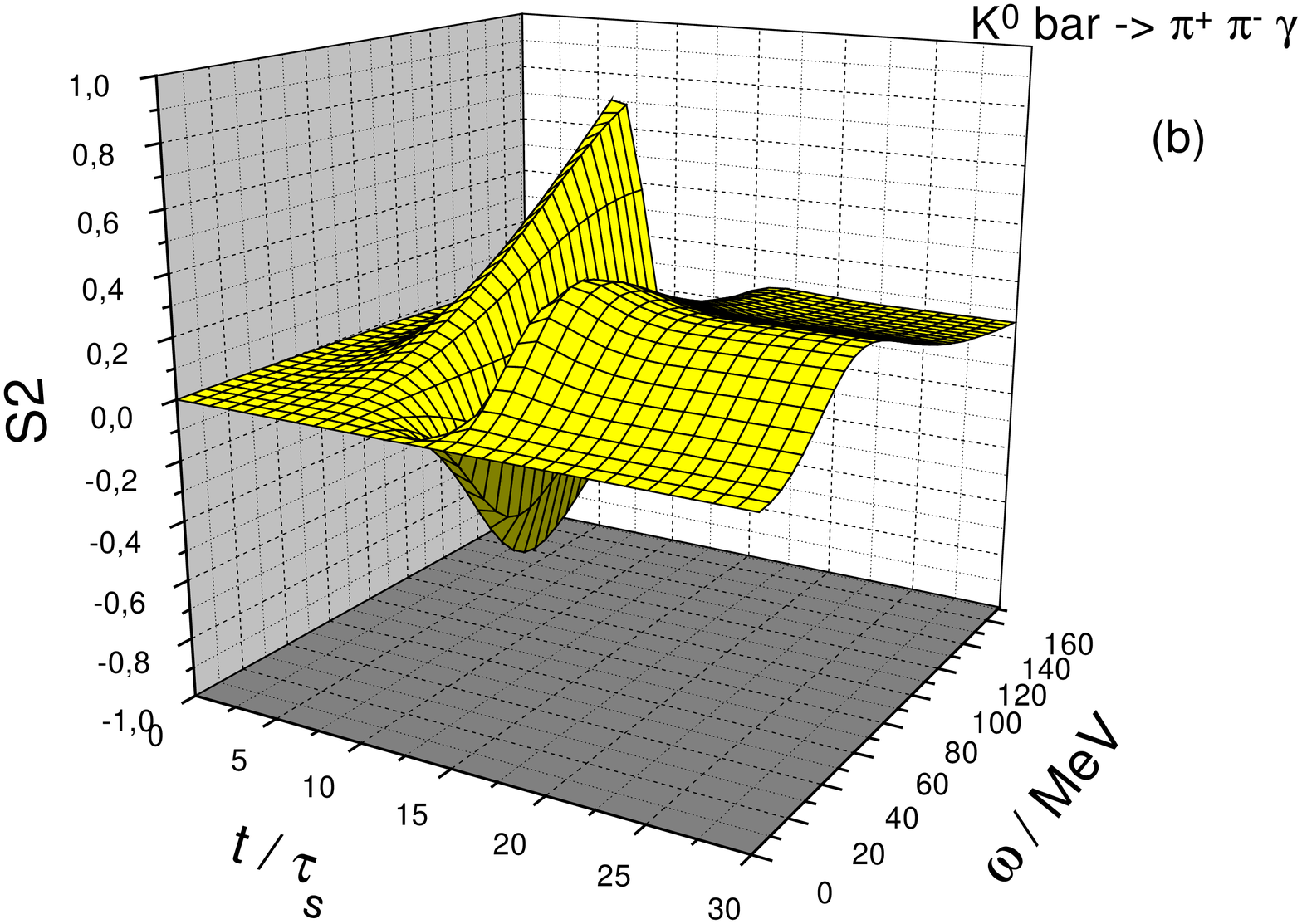}}
}
\vspace{-40pt}
\caption{Components $S_1$ and $S_2$ of the Stokes vector of the photon as a
function of photon energy and time for the decay $\overline{K^0} \to \pi^+ \pi^- \gamma$.}
\label{S1t:S2tb}
\end{figure}

The amplitudes in
Eq.(\ref{AmpK0}) allow us to determine the Stokes vector of the photon at any time
$t$. As an example, Figs.~\ref{S1t:S2ta},~\ref{S1t:S2tb} show the components $S_1(t)$ and $S_2(t)$
as function of energy for an initial $K^0$ or $\overline{K^0}$. It is interesting
to ask how this time dependence would be reflected in the decay spectrum of
$K^0(\overline{K^0}) \to \pi^+ \pi^- e^+ e^-$.

This question has been analysed in~\cite{Sehgal:Leusen00}. In particular, we have
calculated the time-dependent correlation of the $\pi^+ \pi^-$ and $e^+ e^-$ planes,
\begin{equation}
\frac{d\Gamma}{d\phi} \sim 1 - \left( \Sigma_3(t) \cos 2 \phi + \Sigma_1(t)
\sin 2\phi \right),
\end{equation}
and the associated asymmetry ${\cal A}_{\phi}(t)$ defined as in Eq.(\ref{Asymmetry}).

\begin{figure}[htb]
\makebox[75mm]{
\resizebox{75mm}{47mm}
{\includegraphics*[0mm,26mm][280mm,188mm]{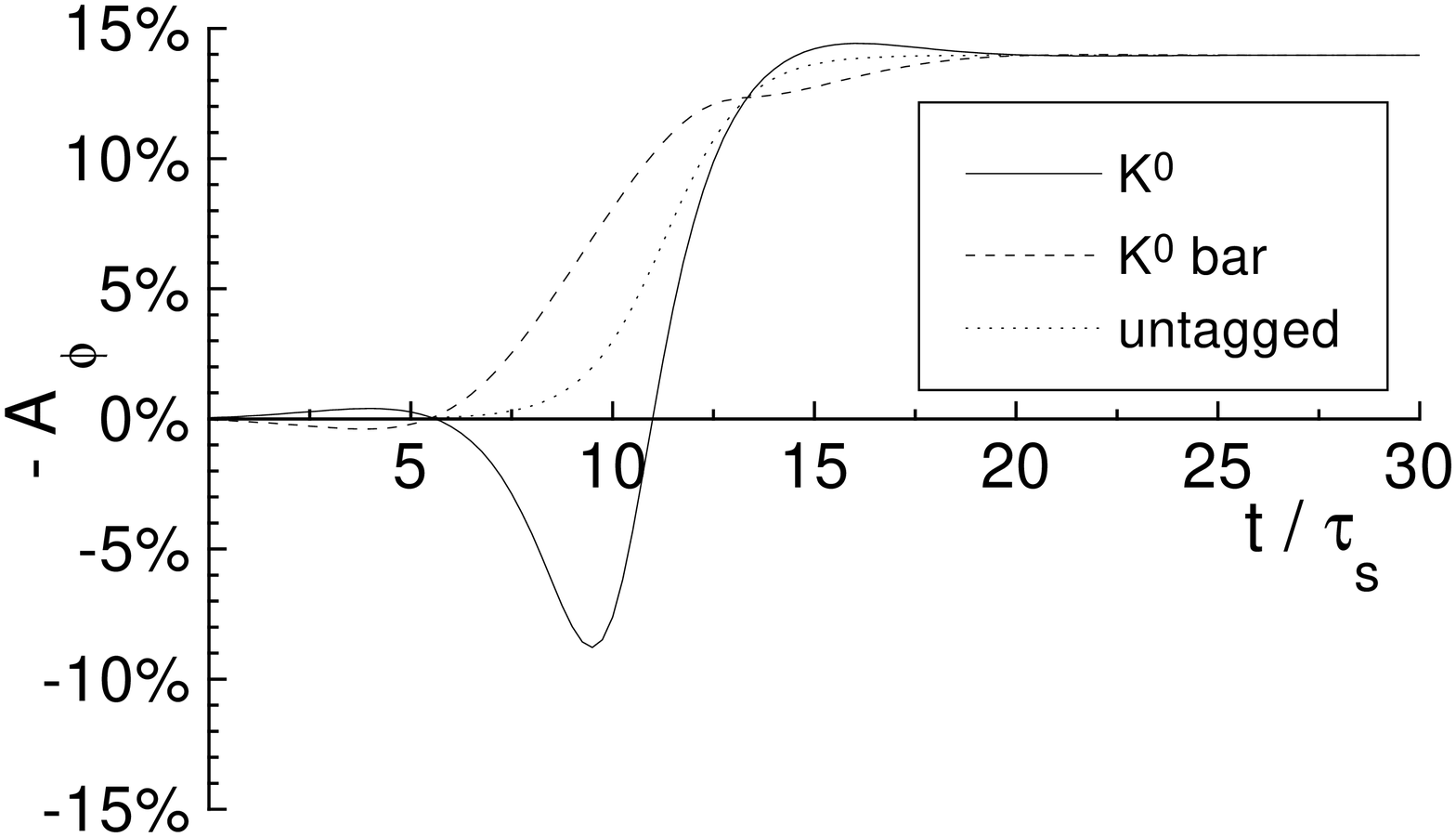}}
}
\vspace{-40pt}
\caption{Time-dependent asymmetry ${\cal A}_{\phi}$ for the decays $K^0$ and
$\overline{K^0} \to \pi^+ \pi^- e^+ e^-$ as well as for an incoherent equal mixture.}
\label{AsyPlot}
\end{figure}

The result is displayed in Fig.~\ref{AsyPlot}, and shows a remarkable time variation,
that differs between $K^0$ and $\overline{K^0}$. As expected (and as measured by
NA48~\cite{NA48}), the asymmetry vanishes at short times, where the $K$ meson
decays essentially as $K_S$ (and the amplitude of $K_S \to \pi^+ \pi^- \gamma$ is
purely electric). At large times the asymmetry approaches the asymptotic value
${\cal A}_{\phi} = - 14 \%$ expected for $K_L \to \pi^+ \pi^- e^+ e^-$. Also
shown in Fig.~\ref{AsyPlot} is the result to be expected if the source of $K$-mesons
is an untagged equal mixture of $K^0$ and $\overline{K^0}$ (such as derived from
$\phi \to K^0\overline{K^0}$ at DA$\Phi$NE). The non-zero value of
${\cal A}_{\phi}(t)$ for such an untagged beam represents a $CP$- and $T$-violating
effect at any decay time.

\section{Acknowledgment}

It is a pleasure to thank Professor M. Savri\'{e} for the invitation to give this talk
and participate in a splendid conference.

\end{document}